**Near-field investigation of a plasmonic-photonic hybrid nanolaser**

Taiping Zhang, Ali Belarouci*, Ségolène Callard*, Cécile Jamois, Xavier Letartre, Céline Chevalier, Pedro Rojo-Romeo, Brice Devif and Pierre Viktorovitch*

*ali.belarouci@ec-lyon.fr;
segolene.callard@ec-lyon.fr;
pierre.viktorovitch@ec-lyon.fr

**The approach to localize light into an ultra small space is being intensively developed to understand the interaction between light and matter in the emerging field of nanophotonics. The plasmonic nanodevices are capable of realizing a confinement of light in sub-wavelength volume [1, 2]. On the other hand, the high index dielectric nanostructures such as photonic crystal (PC) cavities can reach high quality factors (Q-factors) and small mode volumes [3]. Recently, different approaches based on coupling engineering between plasmonic devices and PC cavities, thus resulting in the formation of plasmonic-photonic hybrid nano-devices, have been reported [4 - 9]. Herein, we report an approach of realization and characterization of a novel plasmonic-photonic hybrid nanodevice. The device comprises a plasmonic nano-antenna (NA) and a defect mode based PC cavity. Laser emissions of the devices were demonstrated and the coupling between the NA and PC cavity in different relative positions were investigated via experiment and numerical simulation. Through this study, the coupling sensitivity depend on the polarization of the hybrid device was revealed. It may help us to create better nanophotonic systems for the harnessing of light and the control of light-matter interaction.**

Understanding the interaction between the light and matter in an ultra small space is of utmost importance to build fundamental knowledge in optics. Major efforts have been made to develop techniques for localizing the light into a subwavelength or a nanoscale volume. One route consists of creating defect modes in two-dimensional photonic crystal (2D PC) [10]. 2D PC dielectric slabs have been widely used to form nano or microcavity. The defect is defined by removing several holes from the centre of the array to form a nanocavity. Therefore, light with a certain frequency can be confined in the cavity. Although the optic mode volume of the 2D PC cavity is restricted by the diffraction limit, it can hold a high Q-factor. The interaction and coupling condition between these nanocavities and other photonic nanodevices have attracted a growing interest: for example, nanodiamond with nitrogen vacancy center and quantum dot [11, 12]. In recent years, metallic nanodevices based on localized surface plasmon resonance have received significant attention [13]. The application of these devices has become another facile method to localize the optical field into ultra small volume [14]. The collective electron oscillations at the surface of a metallic nanoparticle yield electromagnetic excitations at the interface of the nanoparticle and the dielectric environment, and localized plasmonic devices have the ability to concentrate electromagnetic energy in an extremely small space. Investigations on the interaction and coupling condition between the PC cavities and plasmonic devices is undergoing [15-17]. These approaches may combine the high Q-factor of PC cavity and the remarkable light confinement ability of plasmonic device and open a new route to harnessing the light in nanoscale volume.

In our previous work, we built a plasmonic-photonic crystal hybrid device that contains a CL7 (missing 7 holes from the centre) PC cavity and a bowtie plasmonic nanoantenna (NA), and successfully demonstrated a laser mode from these hybrid structures. Meanwhile, we found that, under some conditions, the presence of the NA increased the laser threshold of the PC cavity, which means the NA induce losses to the hybrid system [9]. In the work presented in this paper, we characterize the optical field distribution of these hybrid nanodevices. To achieve this goal, a high-resolution optical microscope has been used to get the high resolution optical image by overcoming the diffraction limit. One of the best candidates is scanning near-field optical microscope (SNOM) [18, 19]. Herein we use a SNOM to investigate plasmonic-CL7 PC cavity hybrid nanosystem. We study the near-field optical field distributions of lasing modes of structures in two typical coupling conditions. Our results demonstrate that the resulting mode in this plasmonic-dielectric structures is indeed a hybrid mode. We also analyze the relationship between the optical modes and the relative positions of the plasmonic nano-antenna in the microcavity. The results show that when the NA is oriented parallel to the polarization of the PC cavity, it can localize and enhance the optical field in its gap. Otherwise, when the NA is oriented perpendicular to the polarization of the PC cavity, it cannot localize the optical field. Meanwhile, the numerical simulations of the hybrid devices show that the NA impact the Q-factor of the PC cavity and induce a wavelength movement.

The basic building block of the hybrid nano-device contains a defect PC microcavity and a bowtie nano-antenna (Figure 1a and 1d). The PC microcavity, studied in previous studies [9, 18] consists of a triangular array of cylindrical holes (period 420 nm and hole radius are 90 nm, 100 nm and 110 nm). It is patterned on a thin InP slab (thickness 250 nm) that are positioned on top of a $SiO_2$ substrate. The slab contains four InAsP quantum wells that work as active material to achieve laser emission. The cavity is formed by introducing a linear defect (omitting 7 holes) into the 2D PC. The full structure lateral size is 11 μm × 11 μm. The cavity structure is designed to support several spectrally and spatially distinct optical modes, around 1.5 μm. The computed resonant wavelength λ of the fundamental mode (100 nm holes radius ) is at 1530 nm and the quality factor can be reached to Q = 5800 [18]. To reach reasonably high Q-factors, the two holes on either side of the microcavies are shifted 80 nm outward to achieve an increase by up to 60 percent of the Q-factors compared to unmodified microcavities [18]. Figure 1b shows the optical field distribution of the fundamental mode of the CL7 PC cavity as simulated via 3D-finite-difference time-domain (FDTD). The x, y and z directions of the PC cavity are as shown in the figure and the computational meshes were 10 nm. On the xy plane 30 nm above the cavity, the mode is highly confined into the cavity with the intensity of the mode concentrated principally at the centre of the cavity. The simulation of optical field distribution with Ex and Ey polarizations are shown in the same figure. The optical field in the centre of the cavity along the long axis of the cavity is essentially Ey polarized, while the optical field distributed along the side of the cavity is Ex polarized. From the simulations of the optical field intensity distribution in the xz and yz planes through the centre of the cavity, we can see that the optical field is confined in the InP membrane and that the strongest intensity is found in the middle of the membrane. A scanning near-field optical microscope (SNOM) (Figure 1c i) is employed to investigate the laser emission of the bare cavities [19]. The measurement results of a CL7 cavity is shown in Figure 1c. A far-field pattern of the laser is shown as well (c iii). From the laser spectrum measured in near-field (c ii), one can see the wavelength of the fundamental mode is 1543.00 nm. The measured topography and optical field distribution (c iv and c v) of xy plane shows good agreement with the results of the simulation.

The plasmonic building block of the hybrid nanodevice consists of a gold bowtie NA (figure 1d). The bowtie geometry ($W$ = 140 nm, $L$ = 270 nm, height = 40 nm, feed gap size = 20 nm) has been optimized to hold a plasmonic resonance that matches the spectral position of the PC cavity's photonic mode [9]. To understand the mechanisms of the optical field confinement and enhancement in bowties, theoretical simulations were done to model the resonant field distributions in the NA based on the relationship between the polarizations of the excited light and the directions of the NA. As shown in figure 1e, we set the direction along the symmetric axes of the triangles as //-axis of the NA, while the perpendicular direction as ⊥-axis. The simulations shows that the light with a polarization along the //-axis is concentrated in the centre of the bowtie gap by the NA while the light with a polarization along the ⊥-axis rather exhibits an enhancement along the outside corners of the bowtie. The field intensity enhancement is about 1000 when the bowtie is excited with a polarization along //-axis and 100 along the ⊥-axis direction (Figure 1e). As been explained in our former work, the NA resonant mode spectrum is broad compared with those of the PC cavity mode, allowing for the spectral overlap between the modes of each resonator [9].

The hybrid system was fabricated through a process mainly based on multi-steps of high accurate E-beam lithography which was described in reference [9]. The NAs were located at 2 positions on the backbone of the PC cavity. In case 1, the NA was positioned at the top edge (or bottom edge) of the cavity, and near a hole. The NA's //-axis was parallel to the x-axis of the cavity, as sketched in figure 2a. The SEM image of figure 2b shows our accurate fabrication of this kind of hybrid nanodevices. At this position, the optical field of the fundamental mode of the cavity is polarized along x, which means that the only component of the field is $E_x$ at the NA location Hence the field polarization is parallel to //-axis of the NA. In case 2, the NA was located in the centre of the cavity, and oriented so that its //-axis is still parallel to the x-axis of the cavity (sketch and SEM image in figure 3a and 3b). However, at this position, the optical field of the fundamental mode of the cavity is polarized along the y-axis. As shown by the simulations results in Figure 1, the resonant mode of NA is indeed sensitive to the polarization of the exciting light and the two cases correspond to two different coupling conditions, one being favourable and the other unfavourable. Therefore the behaviour of the hybrid system is expected to be different in the two cases. In particular, we experimentally investigated in each case the field distribution resulting from the coupling between the cavity mode and the NA mode by SNOM.

We started our investigation with the case 1. Figure 2c shows laser spectrum of the fundamental mode of a hybrid structure measured in near-field by our SNOM set-up. The resonant wavelength is 1545.58 nm. It confirms that the hybrid structure sustains a sufficiently high Q-factor to obtain the laser emission. The far-field laser emission pattern of the hybrid structure mode (Figure 2d) is different from the pattern of bare PC cavity mode, which is a clear indication that the presence of the NA has modified the laser mode. The topography and the optical field distribution of the fundamental mode (in xy plane) of the hybrid cavity are shown in Figure 2e and 2f, respectively. The pattern of the intensity distribution is different from the measurement result of the bare PC cavity (without NA): it is a second indication that the NA modifies the mode. A bright spot is observed at the position of the NA, whereas the neighbouring regions are dark, which attests that the intensity of the optical field in the gap is much stronger than in other regions. 3D-FDTD simulations (Figure 2g) reveal that for this structure, the position and orientation of the NA are optimised to couple the NA mode with the Ex polarization of the cavity mode. As the measured optical field map agrees with the computed results, it confirms that the laser emission of the structure is indeed monitored by the hybrid plasmonic-photonic mode. Moreover, it also demonstrates that the NA mode couples to the optical field when it is properly addressed: here, the NA concentrates the optical field in the gap. It may therefore be stated that, in this case, the coupling scheme of the NA and the PC cavity is a "gap coupling". Note that the enhancement is not as high as in the simulation. This may be due to structural fluctuations of the PC cavity and the NA fabrications. This decreases the rate of the enhancement [5]. The simulation results in the xz and yz planes through the gap centre of the NA are also shown. They show that the NA "extract" the light from the PC cavity into its gap. Therefore, the NA localizes the optical field in three dimensional scale. This may reveals that in "gap coupling" mode, the NA and PC cavity hold a strong coupling

Next we consider the structures where the NA are located as in case 2. Figures 3c to 3f shows SNOM measurements of the fundamental mode of a hybrid structure in case 2. The near-field measured laser wavelength is $\lambda$ = 1485.02 nm (Figure 3c). The far-field laser pattern of the structure (Figure 3d) is different from the bare cavity and the hybrid structure in case 1, which indicates another modification of the mode. As described before, the resonance of the NA is sensitive to the polarization of the exciting source. In this case, the orientation of the NA is optimized for a coupling with the Ex component. However, as there is no Ex component at the location of NA, the hybrid cavity cannot localize the optical field in the gap of the NA. Figure 3e shows the topography image of the hybrid cavity and Figure 3f shows the optical field distribution of the fundamental mode. One can see that the four regions of optical field near the corner of the NA are very bright. The signal from the gap of the NA is much weaker, confirming the polarization sensitivity of the hybrid structure. The simulation of the optical field distribution of this type of hybrid structure, shown in Figure 3g indicates that the coupling situation is complex. The optical field is induced next to the four corners of NA. Since the NA may also couple to the optical field Ex near the corners, the latter are bright. In this case, the NA and the PC cavity exhibit a "corners coupling" scheme, in agreement with the theory predictions [5] and the results of simulation. From the simulation in the xz and yz plane through the NA's gap, we can see that the optical field mainly remains in the PC cavity and the field intensity in the NA's gap is much weaker than that in the PC cavity. It means that the "corners coupling" mode is a weak coupling mode. This behaviour is reproducible and has been recorded on several identical structures. The results of the "gap coupling" and "corners coupling" indicate that the coupling between the PC cavity and the NA is sensitive to the polarization matching condition.

Since the NA is a low Q-factor device, it should induce a global decrease of the Q-factor hybrid cavity mode, with respect to the one of the bare cavity. We tried to validate this hypothesis by comparing 3D FDTD simulations of the normalized laser spectra of the fundamental mode of the bare PC-cavity (figure 4a) and those of the hybrid cavities in the cases of "gap coupling" (figure 4b) and "corner coupling" (figure 4c). The spectrum of the bare PC-cavity has a sharp peak at 1551.4 nm with a full width at a half maximum (FWHM) of about 1.1 nm. On the other hand, the spectra of the hybrid cavity in the case of "gap coupling" and "corner coupling" show peaks at 1553.1 nm and 1552.5 nm with a FWHM of 5.9 nm and 2.9 nm, respectively. The broader peaks of the hybrid cavities means the Q-factors of the cavities decrease. The Q-factors of bare cavity, "gap coupling" and "corner coupling" cavity are 1410, 263 and 535, respectively. The simulation results confirm that the NA decrease the Q-factor of the cavity mode. For the condition of "gap coupling", with the efficient NA field localization, the Q-factor decreases more than the "corner coupling". The reason may attribute to the efficient NA field localization yield more scattering and absorption in the metal, as the report of reference 5. Meanwhile, we found that the coupling between the NA and PC cavity induce a wavelength movement. For the "gap coupling", the wavelength has a red shift of 1.7 nm. And for " corner coupling", the red shift is 1.1 nm. The Q-factor decreasing and wavelength movement may signify that the "gap coupling" is stronger than the "corner coupling".

In summary, we report on the design and the demonstration of a novel nano-optical device based on the coupling engineering between a photonic crystal (PC) cavity and an optical nanoantenna (NA). The hybrid device can be fabricated along accurate in large-scale fabrication techniques. A hybrid laser mode can be obtained from this hybrid system. The optical field distribution of the hybrid mode is sensitive to the polarization. In our experimental characterisation, depending on the relative position and orientation between the NA and PC cavity, the hybrid mode exhibits "gap coupling" or "corners coupling" schemes. These results suggest that the coupling condition between the NA mode and the PC cavity mode is sensitive to the polarization matching condition. Moreover, the numerical simulations indicate that the NA induce losses to the hybrid device hence reduce the Q-factor of the PC cavity. And the coupling of the NA and PC cavity yields a mode wavelength movement. This novel system may open new routes for applications in integrated opto-plasmonic devices for quantum information processing, as efficient single photon sources or nanolasers, or as sensing elements for bio-chemical species.

**Methods Summary**

**Device fabrication**

The hybrid structure is formed on an 250 nm thick active high refractive index InP-based membrane. Four InAsP quantum wells (QWs) as active material for light emission in the 1.5 μm range. The membrane is bonded on a $SiO_2$ substrate. The fabrication process including the following 3 steps: defining alignment marks, realizing the photonic crystal and positioning the optical nano-antenna on the backbone of the photonic structure. Ultimate alignment procedures between photonic and plasmonic devices is achieved via the use of metal marks, defined in the first step of the e-beam lithography followed by lift-off process, on which the next layers of lithography are aligned. The second step is the realization of the photonic crystal devices, defined via e-beam lithography and alignment in a positive resist and transferred first into a silica hard mask and then into the InP membrane by two successive reactive ion etching processes. Finally, in a third step the individual metallic nano-antennas are deterministically positioned on the PC cavity by e-beam lithography alignment and writing in a resist double-layer, followed by electron-beam evaporation of Au with Ti adherence layer, and a lift-off process.

**SNOM measurement set-up**

In our case, we use a homemade SNOM configuration designed based on A commercial SNOM head (NT-MDT SMENA) and a commercial microscope (Axio Observer.D1) to work in transmission mode with transparent samples. The structure was pumped by a 50 mW laser diode with a centre wavelength of 780 nm. The diode is pulsed (6.25% of duty cycle) to avoid overheating the structure. The optical signal in near-field is collected by adielectric optical probe and is directed into a diffraction grating monochromater (Jobin Yvon micro-HR,) equipped with micro-slits and an InGaAs thermoelectrically cooled detector. The SNOM set-up has also been conceived to allow recording the far-field image of the optical signal emitted by the sample. The infra-red light emitted by the sample is collected from the back side of the sample by the same objective (63X) which has been used for the excitation. The light is then sent to another microscope output ports equipped with an IR-camera. Therefore, the image of the radiating light at the surface of the sample can be obtained. It should be mentioned here that the Far-field images are not filtrated and the image may contain optical signals at different wavelengths.

**Reference**

[1] S.A. Maier. *Effective mode volume of nanoscale plasmon cavities,* Optical and Quantum Electronics, 38 : 257-267 (2006).

[2] P. Mühlschlegel, H.-J. Eisler, O.J.F. Martin, B. Hecht and D.W. Pohl. *Resonant Optical Antennas,* Science, 308(5728) : 1607-1609 (2005)

[3] Y. Akahane, T. Asano, B.-S. Song, and S. Noda, "*High-Q photonic nanocavity in a*

*two-dimensional photonic crystal,*" Nature **425**(6961), 944–947 (2003).

[4] F. De Angelis, M. Patrini, G. Das, I. Maksymov, M. Galli, L. Businaro, L.C. Andreani and E. Di Fabrizio. *A Hybrid Plasmonic-Photonic Nanodevice for Label-Free Detection of A Few Molecules,* Nano Lett., 8 : 2321-2327 (2008).

[5] M. Barth, S. Schietinger, S. Fischer, J. Becker, N. Nüsse, T. Aichele, B. Löchel, C. Sönnichsen and O. Benson. *Nanoassembled Plasmonic-Photonic Hybrid Cavity for Tailored Light-Matter Coupling,* Nano Lett., 10 : 891-895 (2010)

[6] F. De Angelis, G. Das, P. Candeloro, M. Patrini, M. Galli, A. Bek, M. Lazzarino, I. Maksymov, C. Liberale, L.C. Andreani and E. Di Fabrizio. *Nanoscale chemical mapping using three-dimensional adiabatic compression of surface plasmon polaritons,* Nature Nanotechnology, 5 : 67-72 (2010).

[7] J. Do, K. N. Sediq, K. Deasy, D. M. Coles, J. Rodríguez-fernández, J. Feldmann and D. G. Lidzey. *Photonic crystal nanocavities containing plasmonic nanoparticles assembled using a laser-printing technique,* Adv. Optical Mater., 1 : 946-951 (2013).

[8] T. Zhang, S. Callard, C. Jamois, C. Chevalier, D. Feng and A. Belarouci. *Plasmonic-photonic crystal coupled nanolaser,* Nanotechnology, 25(31) : 315201 (2014).

[9] Y. Yi, T. Asano, Y. Tanaka, B.-S. Song and S. Noda. *Investigation of electric/magnetic local interaction between Si photonic-crystal nanocavities and Au meta-atoms,* Optics Letters, 39(19) : 5701-5704 (2014).

[10] S. Noda, A. Chutinan and M Imada. *Trapping and emission of photons by a single defect in a photonic bandgap structure,* Nature, 407 : 608-610 (2000).

[11] J. Wolters, A. W. Schell, G. Kewes, N. Nüsse, M. Schoengen, H. Döscher, T. Hannappel, B. Löchel, M. Barth and O. Benson. *Enhancement of the Zero phonon line emission from a single nitrogen vacancy center in a nanodiamond via coupling to a photonic crystal cavity,* Appl. Phys. Lett., 97(14) :141108 (2010)

[12] A. Badolato, K. Hennessy, M. Atatüre, J. Dreiser, E. Hu, P. M. Petroff and A. Imamoğlu. *Deterministic coupling of single quantum dots to single nanocavity mmodes,* Science, 308 : 1158-1161 (2005).

[13] 2012 Plasmonics Focus issue Nat. Photonics, 6 : 707-794.

[14] J. A. Schuller, E. S. Barnard, W. Cai, Y. J. Jun, J. S. White and M. L. Brongersma. *Plasmonics for extreme light concentration and manipulation,* Nat. Mater., 9 : 193-204 (2010).

[15] A. Belarouci, T. Benyattou, X. Letartre and P. Viktorovitch. *3D light harnessing based on couping engineering between 1D-2D Photonic Crystal memberanes and metallic nano-antenna,* Optics Express, 18(S3) : A381-A394 (2010).

[16] I.S. Maksymov and A.E. Miroshnichenko. *Active control over nanofocusing with*

*nanorod plasmonic antannas,* Optics Express, 19(7) : 5888-5894 (2011).

[17] A. E. Eter, T. Grosjean, P. Viktorovitch, X. Letartre, T. Benyattou and F. I. Baida. *Huge light-enhancement by coupling a bowtie nano-antenna's plasmonic resonance to a photonic crystal mode,* Optics Express, 22(12) :14464-14472 (2014).

[18] G. Le Gac, A. Rahmani, C. Seassal, E. Picard, E. Hadji, S. Callard. *Tuning of an active photonic crystal cavity by an hybrid silica/silicon near-field probe,* Optics Express, 17(24) : 21672-21679 (2009)

[19] T-P. Vo, A. Rahmani, A. Belarouci, C. Seassal, D. Nedeljkovic and S. Callard. *Near-field and far-field analysis of an azimuthally polarized slow Bloch mode microlaser,* Optics Express 18(26) : 26879-26886 (2010)

**Acknowledgments**

We acknowledge the financial support from CNRS and China Scholarship Council, as well as the technical staff of Nanolyon platform for support and fruitful discussions.

**Author contributions**

T. Z. and A. B. designed the hybrid nanodevice. T. Z. fabricated the nanodevices and did the SNOM measurement. A. B. did the 3D FDTD simulations. S. C. directed the SNOM measurement. C. J. and C. C. directed the e-beam lithography. P. R. R. did the RIE process. B. D. deposited the $SiO_2$ hard mask layer. T. Z., A. B., S. C., C. J., X. L. and P. V. discussed the SNOM measurement results. T. Z. wrote the manuscript. T.Z. and P. V. revised the manuscript. A. B., S. C. and P. V. supervised the project.

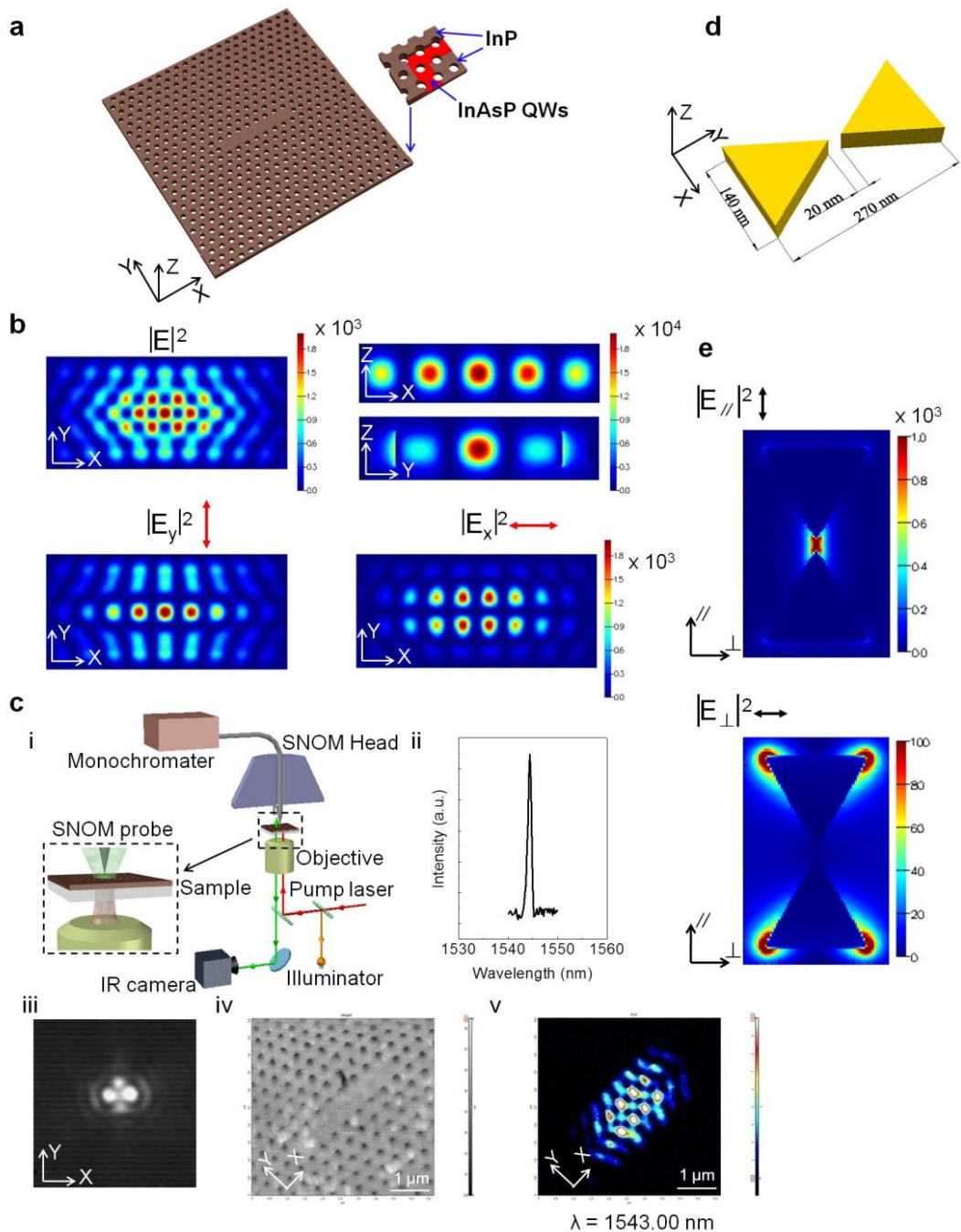

*Figure 1 (a) view of the CL7 PC cavity in the plane of the InP membrane (b) Numerical simulation of optical field distribution of fundamental mode of CL7 PC cavity and profile of the optical field domain with Ex and Ey polarization; (c) SNOM set-up and measured the spectrum and optical distribution of the fundamental mode of CL7 PC cavity; (d) Sketch of bowtie geometry: W=140 nm, L=270 nm, gap=20 nm, height =40 nm. (e) Near-field intensity mapping ($|E/E_0|^2$) of the bowtie antenna with incident polarization in the y-direction and the x-direction.*

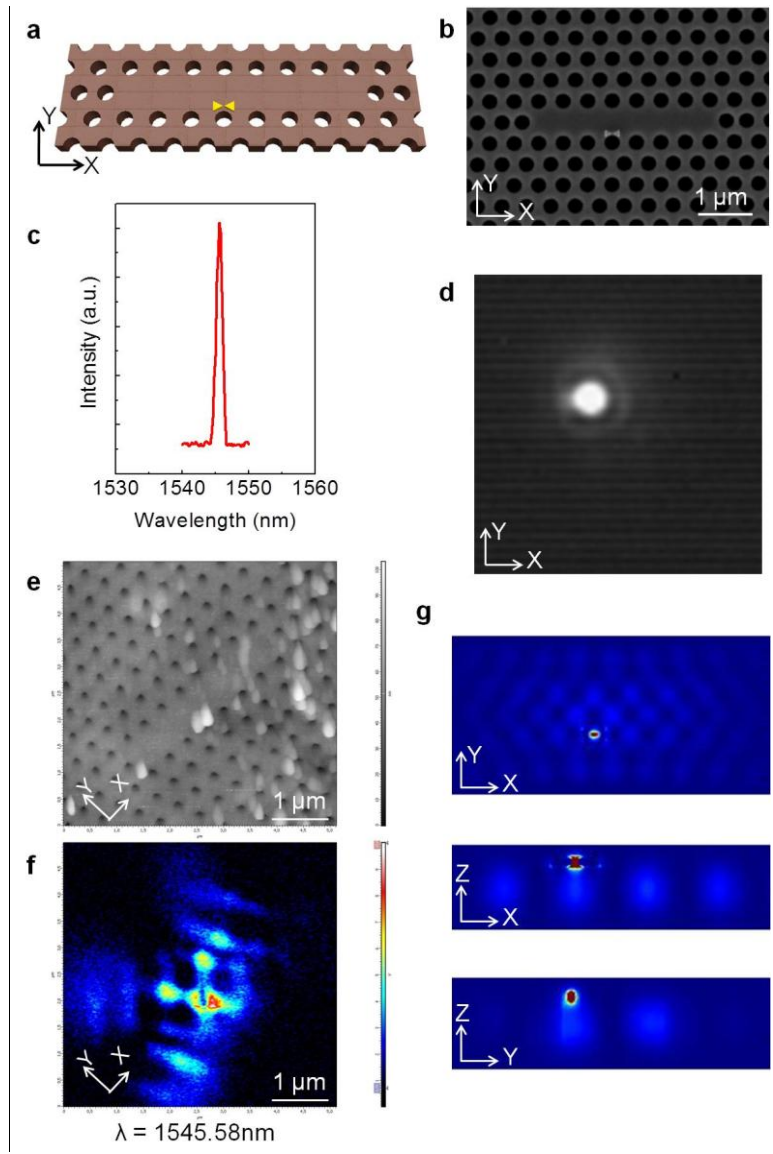

*Figure 2 a, the hybrid cavity of "gap coupling" structure; b, SEM image of the hybrid cavity of "gap coupling" structure; c, Near-field spectrum of the structure; d, Far-field pattern of the structure; e, Topography of the hybrid cavity measured by SNOM; f, Near-field optical field map of the hybrid cavity at fixed wavelength of λ = 1545.58nm; g, simulated optical field map.*

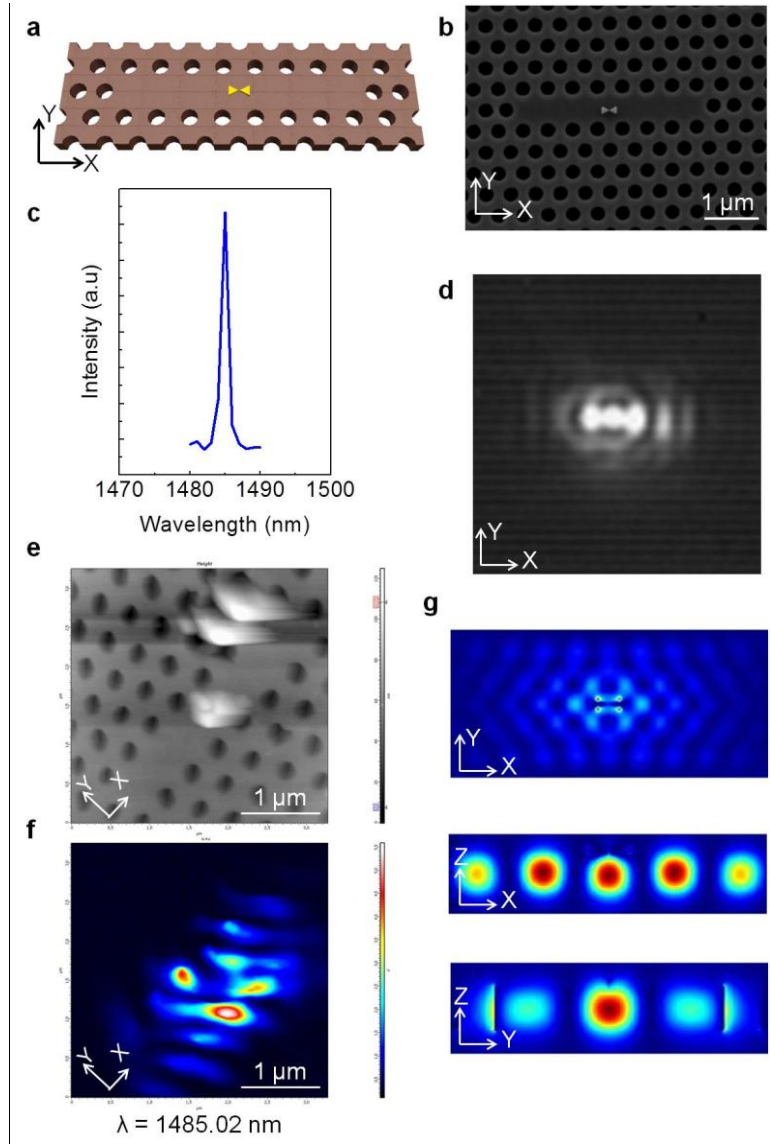

*Figure 3 a, the hybrid cavity of "corners coupling"; b, SEM image of the hybrid cavity; c, Near-field spectrum of the hybrid structure; d, Far-field pattern of the structure; e, SNOM measurement results of topography of the PC cavity; f, SNOM measurement results of optical field maps of the fundamental mode at λ = 1485.02nm; g, numerical simulation of the optical field map of the fundamental mode.*

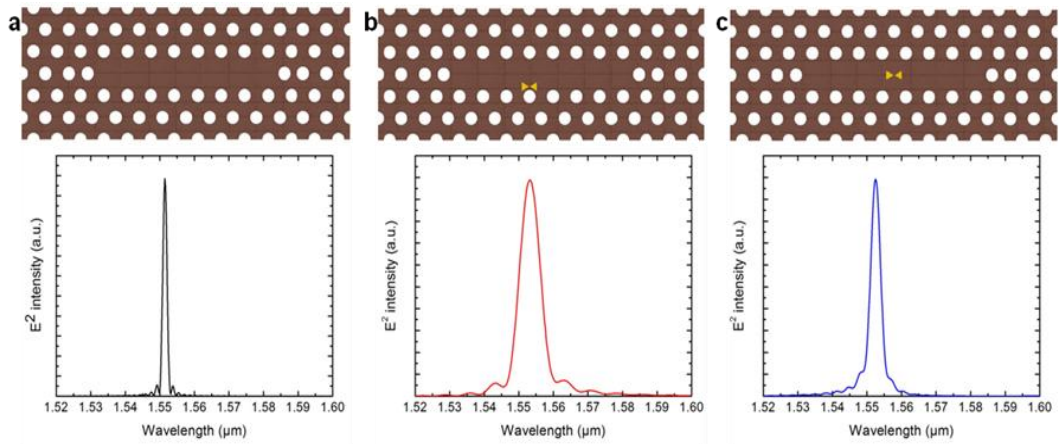

*Figure 4 the normalized spectra of the fundamental modes of: a, the bare PC-cavity, laser wavelength at λ = 1551.4 nm; b, the hybrid cavity of "gap coupling", laser wavelength at λ = 1553.1 nm; c, the hybrid cavity of "corner coupling", laser wavelength at λ = 1552.5 nm.*